\documentclass[aps,manuscript,showpacs,showkeys,superscriptaddress]{revtex4-1}
\usepackage{graphicx}
\usepackage{epsfig}  
\usepackage{epsf}    
\usepackage{dcolumn}
\usepackage{bm}
\usepackage{dcolumn}
\usepackage{textcomp}
\usepackage[tbtags]{amsmath}
\usepackage{amsfonts}
\usepackage{float}
\usepackage{subfig}
\usepackage[]{hyperref}
  \hypersetup{
  unicode=false,          
  pdftoolbar=true,        
  pdfmenubar=true,        
  pdffitwindow=true,     
  pdfstartview={FitH},    
  pdfsubject={Temperature dependent PL study},   
  pdfnewwindow=true,      
  pdfcreator={RevTeX},
  colorlinks=true,       
  linkcolor=red,          
  citecolor=blue,        
  urlcolor=blue,           
  }
\usepackage{hypcap}

\begin{document}

\title{Investigations of temperature-dependent photoluminescence of uncoated and silver-coated CdS 
quantum dots}


\author{P. Ghosh \footnote{Corresponding author}}
\email[Email Address: ]{ghosh.pintu@iitb.ac.in}
\affiliation{Department of Physics, Indian Institute of Technology Bombay, Mumbai 400076, India}

\author{D. Kushavah}
\affiliation{Department of Physics, Indian Institute of Technology Bombay, Mumbai 400076, India}

\author{P. K. Mohapatra}
\affiliation{Department of Physics, Indian Institute of Technology Bombay, Mumbai 400076, India}

\author{P. Vasa}
\affiliation{Department of Physics, Indian Institute of Technology Bombay, Mumbai 400076, India}

\author{K. C. Rustagi}
\affiliation{Department of Physics, Indian Institute of Technology Bombay, Mumbai 400076, India}

\author{B. P. Singh}
\affiliation{Department of Physics, Indian Institute of Technology Bombay, Mumbai 400076, India}
\date{\today}
\begin{abstract}
We report temperature-dependent photoluminescence (PL) studies and PL relaxation dynamics for uncoated 
and silver-coated CdS quantum dots (QDs).
Room-temperature photoluminescence studies indicate that the presence of silver coating on CdS QD samples 
enhances the blue luminescence corresponding to the band-to-band transitions 
by a factor of ten compared to the uncoated samples. 
Furthermore, the decay time measurements using Time Correlated Single Photon Counting technique demonstrate 
F\"{o}rster resonance energy transfer (FRET) between silver-shell and CdS-core in silver-coated CdS QD samples.
A detailed investigation of temperature-dependent PL spectra of uncoated 
CdS QD samples reveals the role of thermally-activated surface-trap states, exciton-LO phonon coupling and 
ionized impurity scattering. 
In case of silver-coated QDs, the temperature-dependent PL peak energy 
corresponding to the band-to-band transitions presents 
a consecutive red-blue-red shift (S-shaped) behavior. 
Whereas, the full width at half maximum (FWHM) shows a successive 
blue-red-blue shift (inverted S-shaped) characteristic with increasing temperature.
\end{abstract}
\maketitle

\section{Introduction}
Semiconductor quantum dots (SQDs) due to their size-tunable energy band-gap and large oscillator strength
have become extensively lucrative for light-emitting diodes (LEDs) \cite{Liangfeng2012, SonDong2012}, 
photovoltaic cells \cite{HuynhScience2002}, 
optically pumped lasers \cite{KlimovScience2000, DuanNature2003}, telecommunications 
\cite{Harrison2009} 
and in biological labels \cite{BruchezScience1998}. 
Coupling of surface plasmon and QD  resonances can open new avenues to further modify their 
spectral characteristics favorably \cite{PompaP}. 
An interplay between plasmonic resonance and the inter-band electronic transitions 
in the SQDs can modify oscillator strength, energy level, PL peak positions and width, and relaxation dynamics. 
In this regard, silver-coated CdS QDs are of special interest because CdS 
quantum dots have high-photoluminescence quantum yield. 
Also, among all metals, silver exhibits 
the most pronounced Mie resonance \cite{Kalyaniwalla1990}.
It is thus important to develop new methods of synthesis and characterization of metal-coated SQDs. 
Mostafavi {\it et al.} reported gamma irradiation method to synthesize highly-monodispersed alloyed 
QDs \cite{Mostafavi2000}. In addition to synthesis and characterization, an in-depth knowledge of the 
radiative and nonradiative relaxation processes in these silver-coated and uncoated SQDs is not only 
intriguing for fundamental physics, but also relevant for future applications in photonics and optoelectronic 
devices. 

Until now, temperature-dependent photoluminescence (TDPL) have been studied for a wide variety of 
SQD samples \cite{Valerini2005, HsuAPL2007}. 
However, to the best of our knowledge, there are hardly 
any such studies in semiconductor / metal core-shell QDs. 
In this paper, we report a detailed investigation of structural determination, 
optical absorption and TDPL studies of such uncoated and silver-coated CdS QDs. 
We find that silver coating on CdS QDs enhances the band-to-band luminescence 
significantly and correspondingly suppresses surface / defect state luminescence. 
A detailed study of the temperature dependence of the PL spectra of uncoated and Ag-coated CdS QDs  
is shown to reveal the role of thermally activated surface-trap states. 
We also demonstrate that the main nonradiative processes are thermal escape of carriers assisted by 
LO phonons and thermally activated ionized impurity scattering. 
\section{Experimental}
For the preparation of CdS nanoparticles, 100 ml of $2.5\times10^{-4}$ M 
cadmium sulphate ($CdSO_4$) solution and 100 ml of $7.5\times10^{-2}$ M 
2-mercaptoethanol ($HOCH_2CH_2SH$) were added to sodium phosphate buffer solution ($pH=7.6$). 
Thereafter, a $Co_{60}$ $\gamma$-ray source with 2.65 $kGy/h$ dose rate was used to irradiate the 
resultant solution.
All the reagents are of analytical grade (Sigma-Aldrich) and are used without further purification. 
Mili-Q water (18 M$\Omega$ cm$^{-1}$) was used to prepare solutions. 
Since the growth of clusters is highly dependent on any type of nuclei, 
the vessels were cleaned with concentrated $HNO_3+ HCl$ solution in order to remove all contaminations. 
The solutions were isolated from air and then de-aerated by bubbling nitrogen.  
The steady state irradiation of the solution results in the continuous 
release of radical species.
Finally CdS molecules are formed through the reaction, 
\begin{equation}
 Cd^{2+}+SH^-\longrightarrow CdS + H^+.
\end{equation}
Consequently, CdS molecules coalesce to form CdS clusters in solution.
Eight samples were prepared varying the dose of gamma irradiation from 0.2 kGy to 7.5 kGy. 
These samples are named as CdS1, CdS2 and so on. 
All samples are tested for any residual radioactivity and are used for further experiments only when 
they were found free of any radioactivity.
\subsection{Deposition of Silver on CdS Nanoparticles}
Silver was deposited on two CdS QD samples namely CdS2 and CdS8 which 
were synthesized with 0.3 and 7.5 kGy dose. 
These two samples after depositing silver are named as CdS2@Ag and CdS8@Ag respectively.
The deposition of silver on CdS nanoparticles was done in three steps. 
First, excess of ionic compounds were removed from CdS 
solution by amberlite treatment and then 10 ml of 0.5 gm wt$\%$ 
polyvinyl sulfate polymer (PVS) was added to avoid 
the coalescence of the CdS nano particles. 
Second, 5 ml of $2\times10^{-4}$ M silver nitrate 
solution was slowly injected under rigorous stirring. 
5 ml of $2\times10^{-4}$ M formaldehyde was added to the CdS solution. 
Formaldehyde was used as the reducing agent.
When silver ions are reduced by formaldehyde 
in presence of CdS particles, the reduction occurs 
instantaneously and individual (Ag)$_n$ particles are 
not produced \cite{Rocco2004}.\\
For structural characterization, high-resolution 
transmission electron microscopy (HRTEM) was used.
Spectroscopic studies such as UV-Vis absorption and 
PL were performed on these colloidal solution of uncoated 
and silver-coated CdS QD samples. 
Temperature-dependent PL spectroscopy was done on 
solid thin films prepared by spin coating technique. 
Clean silicon wafers were used as substrates as they do not luminesce.
A closed cycle He refrigerator was used for cooling 
the sample to 10 K. He-Cd laser (wavelength: 325 nm) 
of power 25 mW was used as the excitation source. 
PL spectra were recorded using a spectrometer with 
CCD camera, cooled to 210 K.
\section{Results and Discussion}
\subsection{Structural characterization}
\begin{figure}[H]
\centering
  \includegraphics[height=0.23\textheight,keepaspectratio]{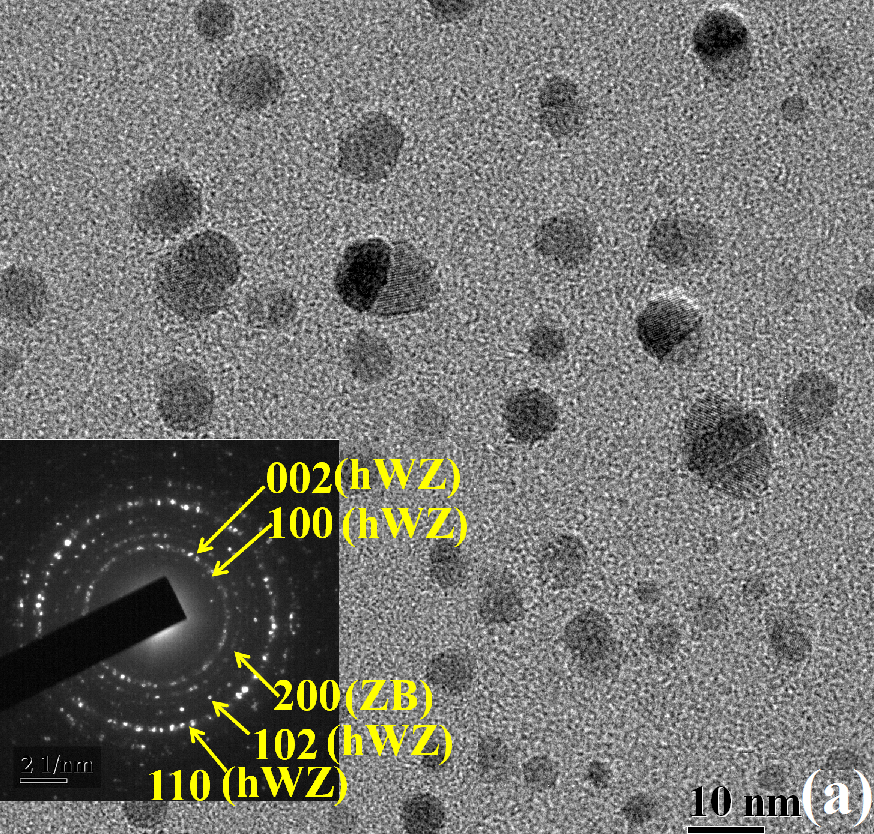}
  \includegraphics[height=0.23\textheight,keepaspectratio]{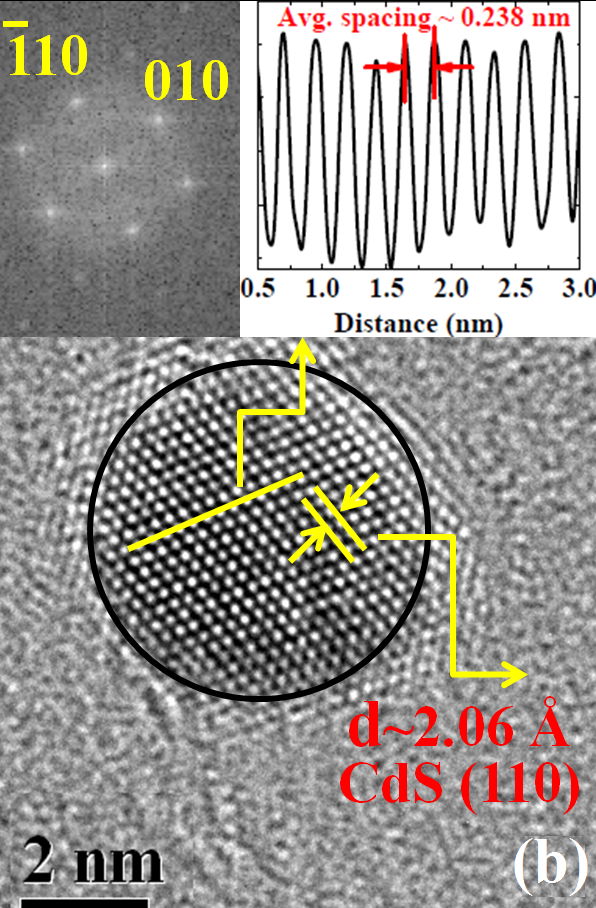} \\
  \includegraphics[height=0.21\textheight,keepaspectratio]{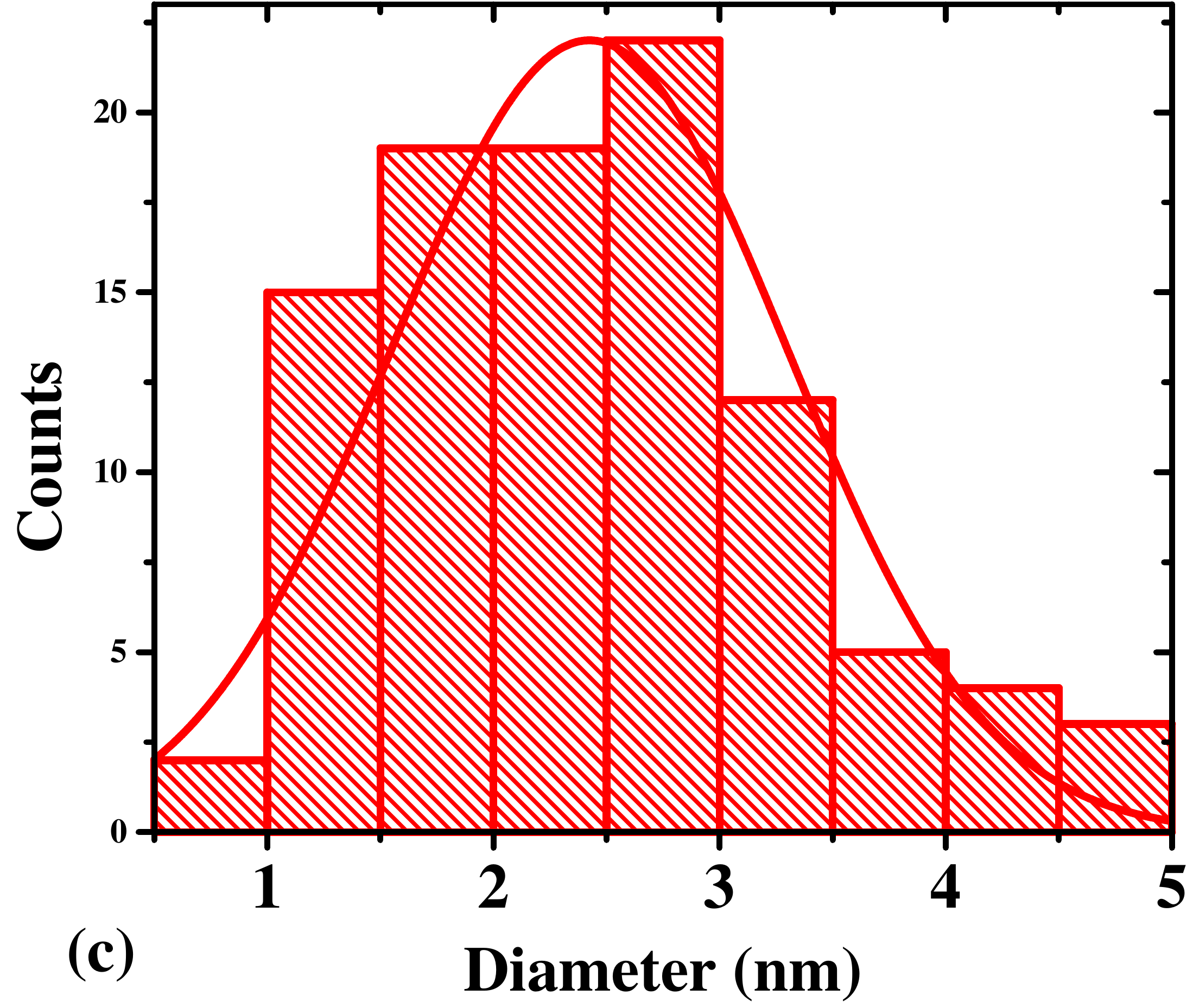}
  \includegraphics[height=0.21\textheight,keepaspectratio]{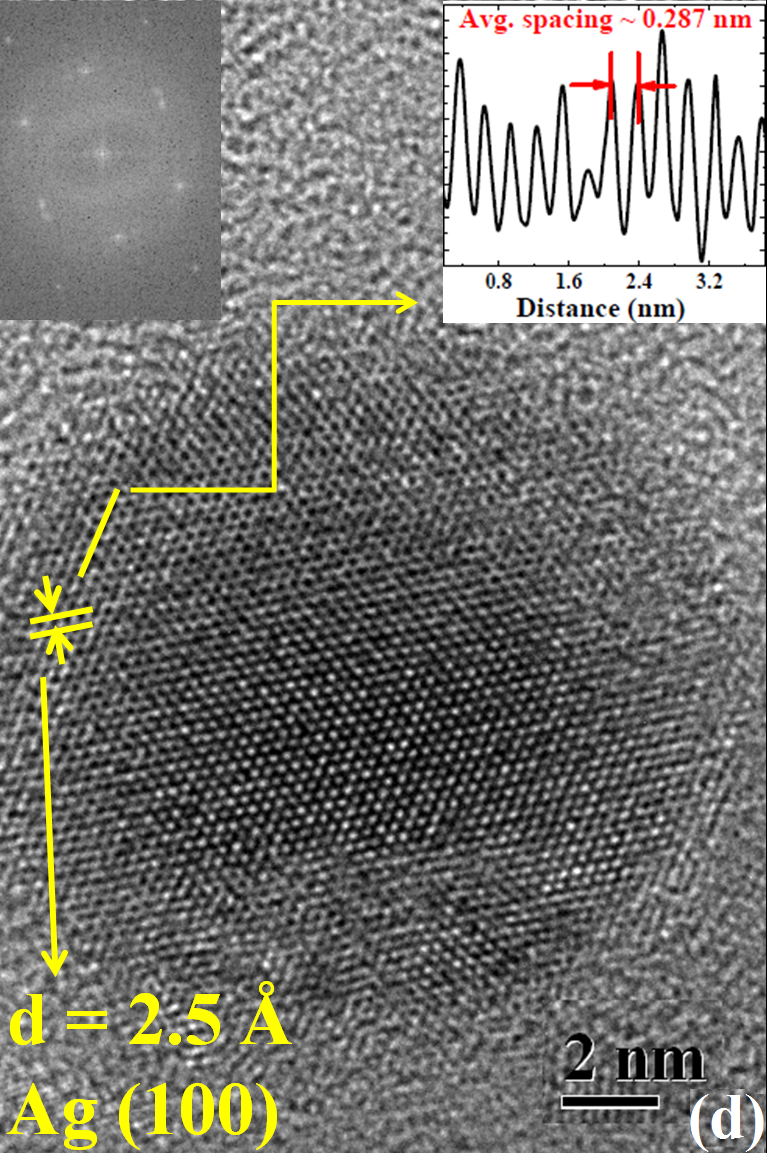}  
  \caption[Representative HRTEM images of silver-coated CdS QDs (CdS2@Ag).]
  {\label{TEM_CdS_CdS@Ag} \footnotesize{
  (a) Representative High-resolution TEM image of silver-coated CdS QDs (CdS2@Ag). 
  (b) HRTEM image on an individual silver-coated QD. 
  The left side inset shows the FFT of this Silver-coated CdS QDs. 
  The right side inset shows the average lattice point spacing.
  (c) Histogram of particle size distribution. 
  (d) HRTEM image of an individual silver-coated CdS QD.}}
\end{figure}

Representative high-resolution transmission electron microscopy (HRTEM)
images of uncoated CdS QDs (CdS2) are shown in Fig. \ref{TEM_CdS_CdS@Ag} (a). 
The inset shows the selected area electron diffraction (SAED) of the QDs.
The lattice spacing ($d$) between adjacent (hkl) 
lattice planes are obtained from the diameter of the diffraction rings. 
Deploying these $d$ values, lattice parameters a and c are obtained. 
Comparing these values with 
literature values for hexagonal-wurtzite (hWZ) and cubic zinc-blende (ZB) crystal structures, we 
deduce information about the crystal structure of these QDs (see Table \ref{d_hkl}).
\begin{table}[H]
\centering
\begin{tabular}{c c c c c c c c}
\hline
\hline
Avg. radius& $d$ & $d$ & $h$ & $k$ & $l$ & Crystal\\ 
$r$ (nm$^{-1}$)& $(= 1/r)$ ($A^o$)& Literature value ($A^o$)& & & &type\\ \hline \hline
2.810 & 3.559 & 3.5688 [PDF 80-0006] & 1 & 0 & 0 & hexagonal\\ 
3.005 & 3.333 & 3.3410 [PDF 80-0006] & 0 & 0 & 2 & hexagonal\\ 
3.448 & 2.900 & 2.9055 [PDF 80-0019] & 2 & 0 & 0 & cubic\\
4.121 & 2.430 & 2.4390 [PDF 80-0006] & 1 & 0 & 2 & hexagonal\\
4.906 & 2.038 & 2.0605 [PDF 80-0006] & 1 & 1 & 0 & hexagonal\\ \hline \hline 
\end{tabular}
\caption{\label{d_hkl}\footnotesize{Table showing spacing d between adjacent (hkl) lattice planes. 
Vertical and horizontal radii, measured from the diffraction rings, are averaged to obtain $r$.}}
\end{table}
In conclusion, the analysis of SAED image indicates the 
presence of planes corresponding to hexagonal-wurtzite (hWZ) and cubic zinc-blende (ZB) crystal structures in 
case of these CdS QDs. It has already been reported for CdS QDs, 
in 4-6 nm range, that the crystal structure phase changes 
from bulk hexagonal-wurtzite type to cubic zinc-blende structure \cite{Ayyub2000}.
The crystal structure has been found to be exclusively dependent on the size of the QD.
For QDs with diameter less than 4 nm, the crystal structure has been reported to be cubic zinc-blende structure;
whereas that with diameter greater than 6 nm, the structure has been reported to be 
hexagonal-wurtzite structure. For the QDs with diameter in the range of 4-6 nm, both the
cubic and hexagonal phase co-exist sharing their close-packed planes at the boundary \cite{Ricolleau1998}. 
Fig. \ref{TEM_CdS_CdS@Ag} (b) and (c) show the 
HRTEM images of individual CdS QDs of diameter 5 nm and 3 nm respectively.
The HRTEM lattice image of individual CdS QDs show the well-spaced lattice 
fringes indicating a single crystal structure of CdS QDs with high crystalline quality. 
The left side inset shows the FFT of the circled region and the right side inset shows the average lattice point 
spacing. This lattice point spacing (L) makes an angle $30^o$ with the lattice plane spacing ($d$). 
Therefore, the lattice plane spacing is obtained using the relation: $d_{hkl}=L \cos(30)$.
The crystal plane spacing comes out to be 2.06 $\mathring{A}$,
which correspond to crystal plane spacing of CdS (220) zinc-blende single crystal or 
CdS (110) of hexagonal-wurtzite single crystal. 
Fig. \ref{TEM_CdS_CdS@Ag} (d) shows an individual silver-coated QD. 
A close look at this image reveals that each particle consist of 
a darker center part compared to the edge. This color contrast is attributed to the silver-coating on CdS QDs.
The left side inset shows the FFT of this Silver-coated CdS QDs. 
The right side inset shows the average lattice point spacing. 
The crystal plane spacing comes out to be 2.5 $\mathring{A}$,
which correspond to crystal plane spacing of Ag (100) hexagonal-wurtzite single crystal [PDF 87-0598].
\subsection{Absorption spectroscopy}
\begin{figure}[H]
\centering
  \includegraphics[height=0.25\textheight,keepaspectratio]{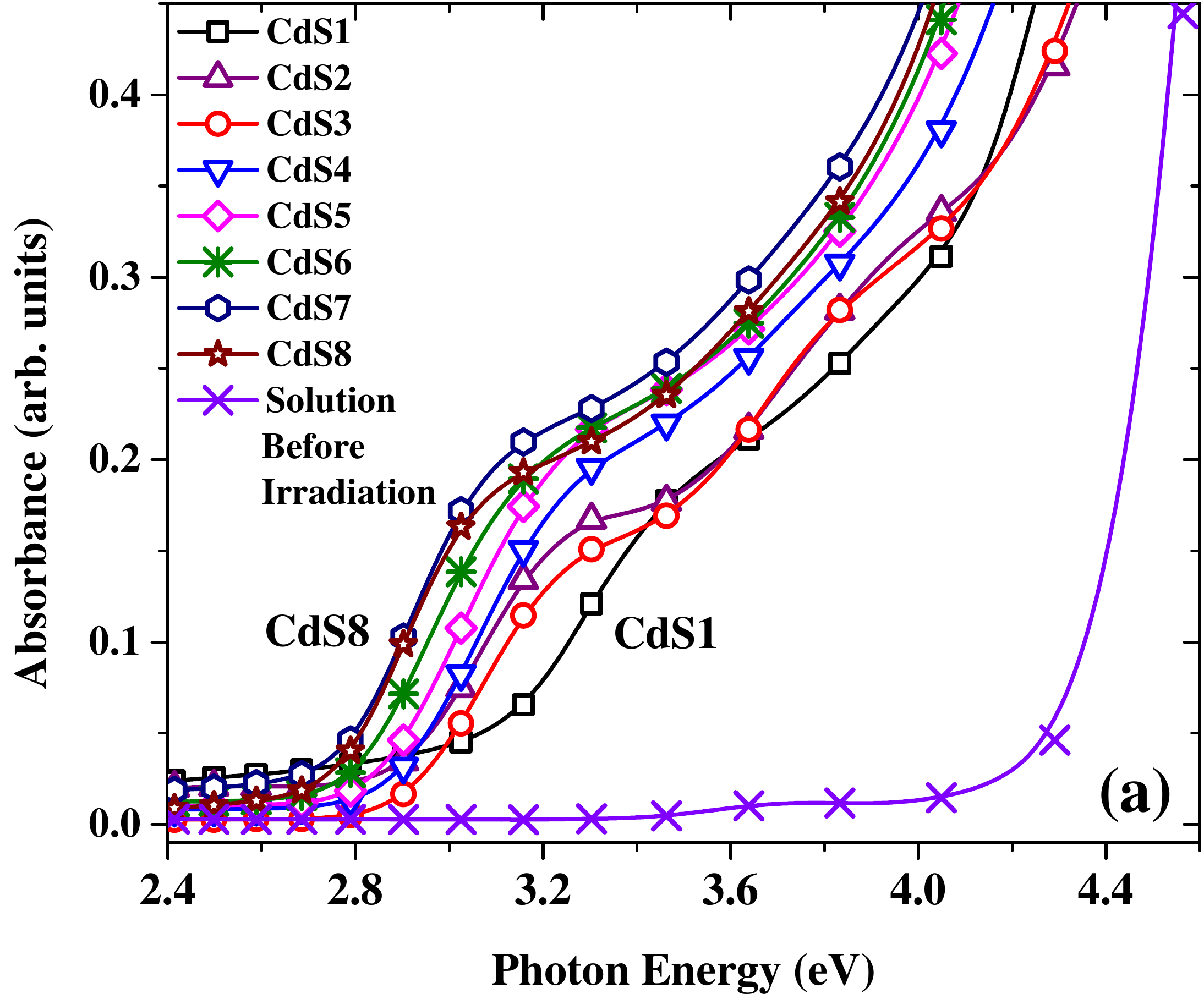}
  \includegraphics[height=0.26\textheight,keepaspectratio]{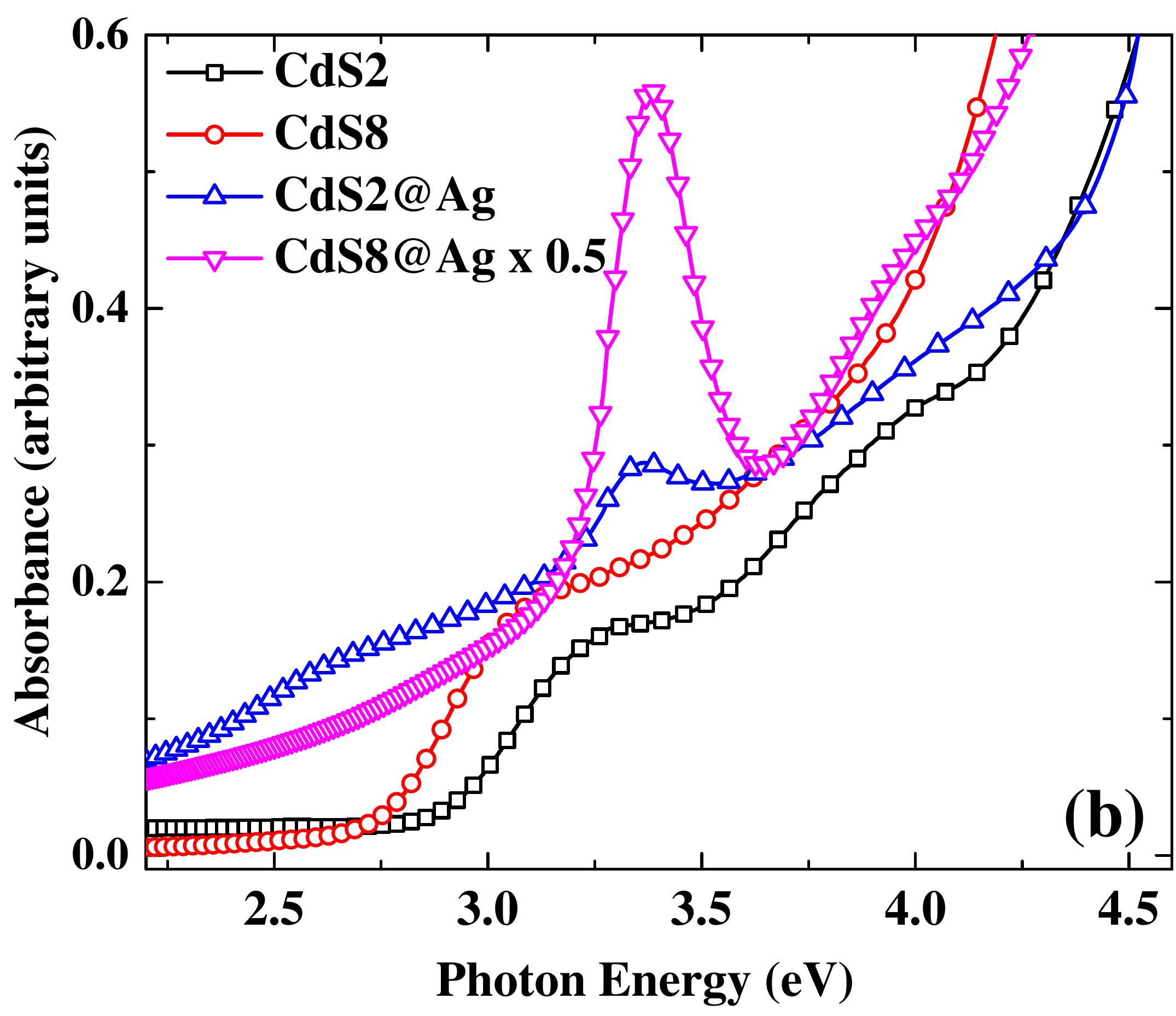}
\caption[Absorption spectra of colloidal solution of uncoated silver-coated CdS QDs.]
{\label{abs_all} \footnotesize{Left panel shows 
absorption spectra of colloidal solution of uncoated CdS QD samples. 
The spectrum for the solution before $\gamma$-irradiation 
is shown for comparison. Right panel shows the comparison of absorption 
spectra of colloidal solutions of uncoated and silver-coated CdS QDs.}}
\end{figure}

A LAMBDA 950 UV/Vis/NIR spectrophotometer (PerkinElmer) is used to record absorption spectra of 
the CdS QD samples.
The UV-Vis absorption spectra of colloidal solutions, 
before and after $\gamma$-irradiation with different doses, 
are shown in Fig. \ref{abs_all} (a). 
The absorption spectra are obtained towards analyzing the 
quantum-size effect of the CdS nanoparticles.
The peak positions of the absorption bands for 
these SQD samples appear in the range 2.9 eV to 3.3 eV. 
While, the band-gap of bulk CdS is 2.42 eV. 
This large blue shift of the band edge is due to strong quantum confinement effect 
in the smaller CdS QDs of radius less than exciton 
Bohr-radius $a_B$ (= 2.8 nm for CdS \cite{Gaponenko1998}).
In case of lower doses (0.2 kGy - 0.4 kGy), 
we observe multiple absorption peaks which appear due to overlap of higher-order 
excited states corresponding to different particle size QDs.
The corresponding band gap for each sample is determined from the $\lambda_{1/2}$ position 
of the excitonic absorption feature. Further, the size of QDs are obtained 
from data reported in Nair ${\it et al.}$ \cite{SelvaNair1992}, 
where they calculate the QD diameter numerically using 
conduction-valence-band model (CVBM) for CdS QD.
Measured band gap and average QD diameter for different 
samples corresponding to different $\gamma$-irradiation dose are shown in Table \ref{table_band_gap}. 
\begin{table}[H]
\begin{center}
\begin{tabular}{c c c c c}
\hline
\textbf{Sample} & \textbf{Dose} 
& \textbf{Bandgap} & \textbf{Diameter}  \\ 
& \textbf{(kGy)}   & \textbf{(eV)} & \textbf{(nm)} \\ \hline
CdS1 & 0.2  & 3.25  & 3.3 \\ \hline
CdS2 & 0.3  & 3.07  & 4.2 \\ \hline
CdS3 & 0.4  & 3.05  & 4.0 \\ \hline
CdS4 & 1.5  & 3.05  & 4.2 \\ \hline
CdS5 & 3.0  & 3.02  & 4.4 \\ \hline
CdS6 & 4.5  & 2.95  & 4.8 \\ \hline
CdS7 & 6.4  & 2.90  & 5.5 \\ \hline
CdS8 & 7.5  & 2.90  & 5.5 \\ \hline
\end{tabular}
\caption{\label{table_band_gap} \footnotesize{Measured band gap and calculated average QD diameter for different 
samples corresponding to different $\gamma$-ray dose.}}
\end{center}
\end{table}
The calculated particle size values are in good 
agreement with the HRTEM results.
With increasing dose, the spectra shift towards red, 
which suggests the formation of larger particles. 
For the solution which was not irradiated, 
an absorption band was observed around 4.5 eV. 
This was due to the formation of complexes 
between $Cd^{2+}$ and $HOCH_2CH_2SH$ \cite{Mostafavi2000}.
In case of silver-coated CdS QDs, absorption 
spectra show a peak at wavelength around 367 nm (3.4 eV) 
as is observed in Fig. \ref{abs_all} (b). 
This absorption peak corresponds to the 
coupling of localized surface plasmon with excitons of CdS-core \cite{Ghosh2015}.
No strong absorption band was observed in the 
380 nm - 400 nm region due to the development 
of pure silver QDs \cite{Rocco2004}. 
\subsection{Photoluminescence spectroscopy}
\begin{figure}[H]
\centering
  \includegraphics[height=0.26\textheight,keepaspectratio]{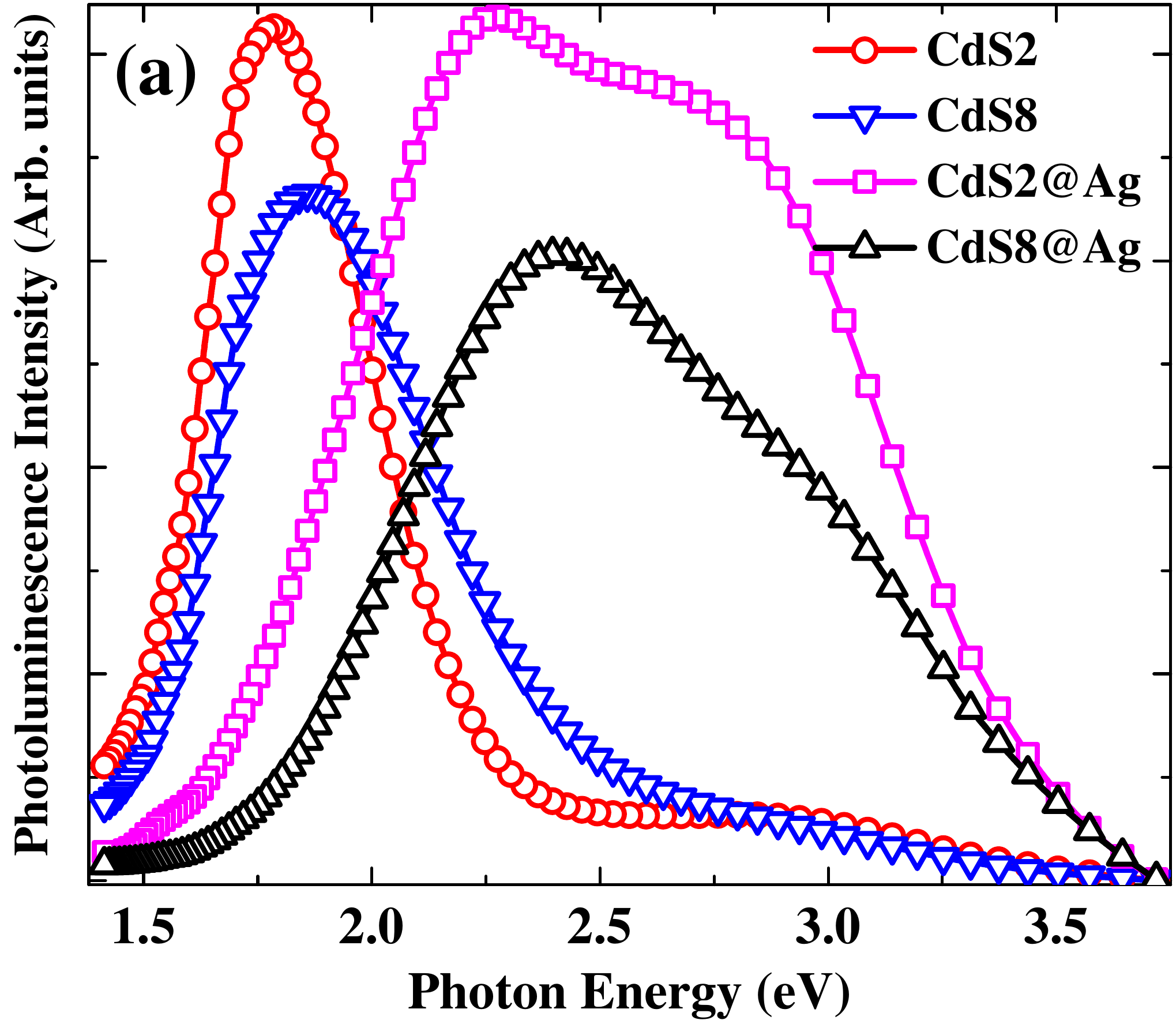}
  \includegraphics[height=0.26\textheight,keepaspectratio]{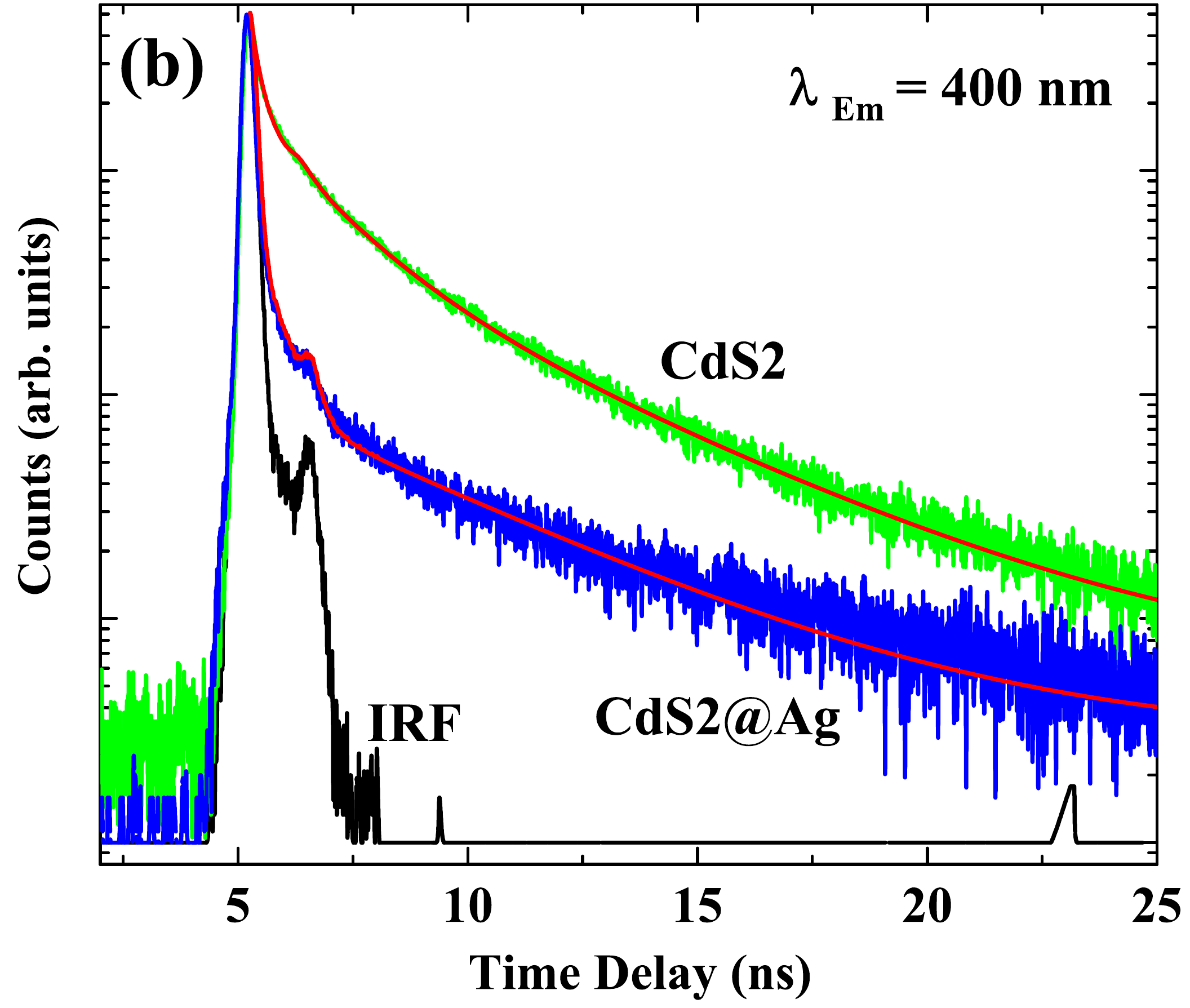}
\caption{\label{RTPL_TCSPC_ALL} \footnotesize{Photoluminescence spectra 
of the colloidal solution of uncoated (CdS2 and CdS8) and silver-coated (CdS2@Ag and CdS8@Ag) samples.}}
\end{figure}

Fig. \ref{RTPL_TCSPC_ALL} (a) shows representative PL spectra 
of the colloidal solution of uncoated (CdS2 and CdS8) and silver-coated (CdS2@Ag and CdS8@Ag) samples.
All the spectra of colloidal solutions of 
uncoated CdS nanoparticles show two broad PL bands. 
The band at higher energy comes in the energy 
range 2.6 eV to 3.4 eV (blue luminescence), whereas 
the band at lower energy comes in the energy 
range 1.4 eV to 2.4 eV (red luminescence). 
The PL peak at higher energy is due to the band 
to band transitions, while the peak in lower 
energy range is due to defect state transitions. 
For uncoated CdS QDs, the PL peak intensity due to 
the defect state transitions is much stronger than 
the one due to band-to-band transitions. 
The defect states arise due to presence of impurities, 
cadmium vacancies, sulfur vacancies or from 
the surface-dangling bonds. 
In case of very small particles, the surface to 
volume ratio is very high. 
With decreasing particle size, this ratio gets 
enhanced and many surface electronic states get 
introduced to the system.
As a result, with decreasing particle size, 
defect state transitions dominate over band-to-band transitions. 
For uncoated CdS samples, the PL peak intensity of 
red luminescence is about ten times higher than that 
of the blue luminescence. Whereas, in case of 
silver-coated CdS nanoparticle solutions, the strength 
of the blue and red luminescence are of the same order.
For silver-coated CdS nanoparticles, the thin layer of 
silver on top of the semiconductor core, effectively 
passivates the surface defect states. 
As a result, the PL peak intensity, due to 
the defect state transitions, goes down with respect 
to the band-to-band transitions. \\
Towards understanding the reason for change of photoluminescence profile in 
presence of silver coating, TCSPC experiments have been performed on these uncoated 
and silver-coated CdS QD samples.
A picosecond laser of wavelength 375 nm is used
as excitation source and the emission is monitored at 400 nm, which is the PL peak position corresponding to the 
band-to-band transitions. 
Full width at half maximum (FWHM) of the instrument response function (IRF) 
is 250 ps. 
The measured TCSPC decay profiles for colloidal solutions of CdS2 
and CdS2@Ag QD samples are shown in Fig. \ref{RTPL_TCSPC_ALL} (b). 
The decay profiles for both the samples could be fitted with three exponential decay function. 
The decay lifetime corresponding to the CdS2 QD sample are $\tau_1 = 197$ ps (44.73), 
$\tau_2 = 1.466$ ns (26.70) and $\tau_3 = 4.497$ ns (28.57). 
Whereas, these values for CdS2@Ag sample come out to be $\tau_1 = 17$ ps (75.02), 
$\tau_2 = 376$ ps (8.38) and $\tau_3 = 4.501$ ns (16.60). 
The number in the in first bracket gives 
the fractional intensity.
It shows that the
faster decay lifetime components corresponding to the 
band-edge transitions become even more fast and the 
amplitude of fractional intensity become dominant in
case of silver-coated sample compared to uncoated sample. 
The presence of metal surface near luminating SQDs may change the spectral profile of the SQDs due to 
F\"{o}rster resonances energy transfer (FRET) from plasmon of 
metal to exciton of semiconductor or vice versa 
\cite{HosokiPhysRevLett2008, Lakowicz2005, Chance2007, Kummerlen1993}. 
In the absence of any quenching or enhancing processes, the lifetime ($\tau_0$) and quantum yield ($Q_0$) 
of a SQD are determined by the rate of radiative ($\Gamma$) 
and nonradiative decay to the ground state ($k_{nr}$). 
The quantum yield and lifetime are given by \cite{Schaufele2005}
$Q_0=\Gamma[\Gamma+k_{nr}]^{-1}$, and $\tau_0=[\Gamma+k_{nr}]^{-1}$. 
According to the above equation, low quantum yield from SQDs implies $k_{nr}>>\Gamma$. 
However, the presence of a metallic surface modifies 
the radiative decay rate of an excited SQD. Considering the radiative rate increases to be $\Gamma_m$, 
the quantum yield ($Q_m$) and lifetime ($\tau_m$) of the SQD become
$Q_m=[\Gamma+\Gamma_m][\Gamma+\Gamma_m+k_{nr}]^{-1}$, and $\tau_m=[\Gamma+\Gamma_m+k_{nr}]^{-1}$, 
respectively.
In principle, $\Gamma_m$ can have values from $-\Gamma$ to + $\infty$. 
As the value of $\Gamma_m$ increases, the lifetime ($\tau_m$) decreases and 
the quantum yield ($Q_m$) increases. 
A careful examination of the above equation reveals that high
quantum yields can be obtained whenever $\Gamma_m$ is comparable to $k_{nr}$. 
Therefore, it can be concluded that the decrease in decay lifetime in case of silver-coated CdS QDs 
is due to resonance energy transfer between plasmon of silver-shell and excitons of CdS-core.
\subsubsection{TDPL of uncoated CdS QDs}
\begin{figure}[H]
\centering
  \includegraphics[height=0.27\textheight]{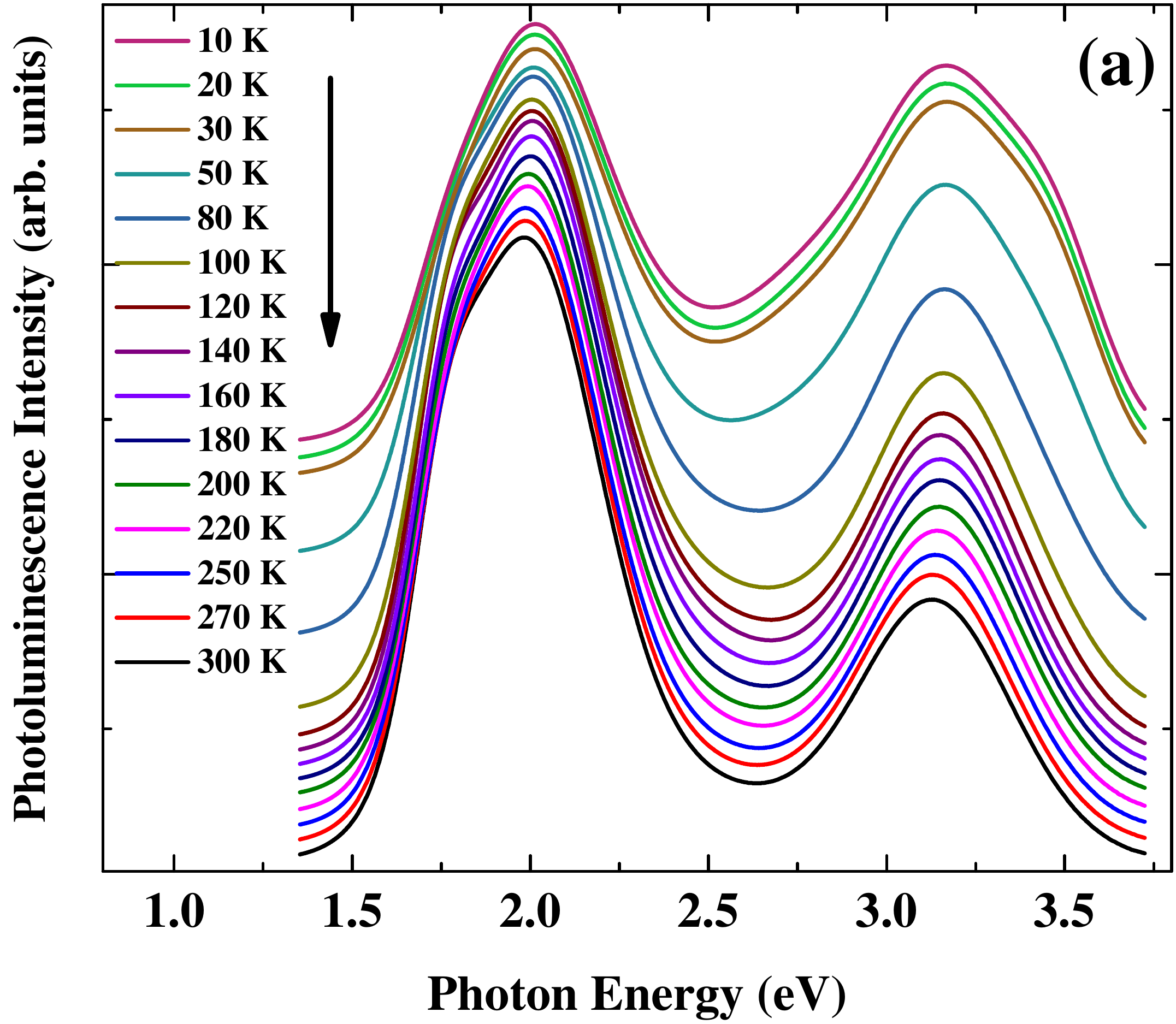}
  \includegraphics[height=0.27\textheight]{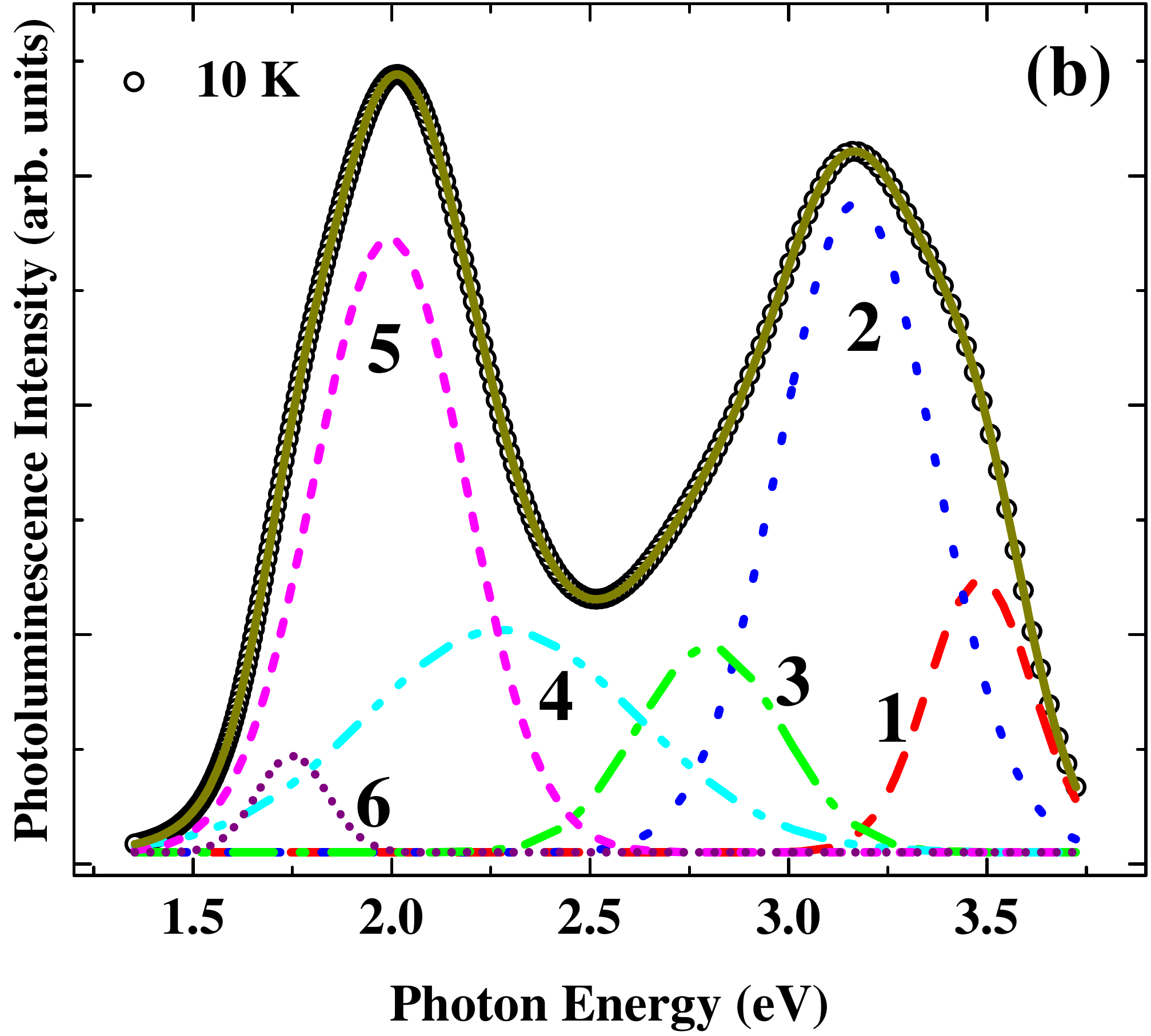}
\caption{\label{TDPL_CdS2} \footnotesize{ (a) Measured temperature-dependent PL spectra of the solid thin film 
of uncoated CdS QDs (CdS2). The curves are shifted along Y-axis for clarity. 
(b) Fitting of the PL spectrum at 10 K with six best-fit Gaussians (dot-dashed lines).}}
\end{figure}

\begin{figure}[H]
\centering
  \includegraphics[height=0.27\textheight]{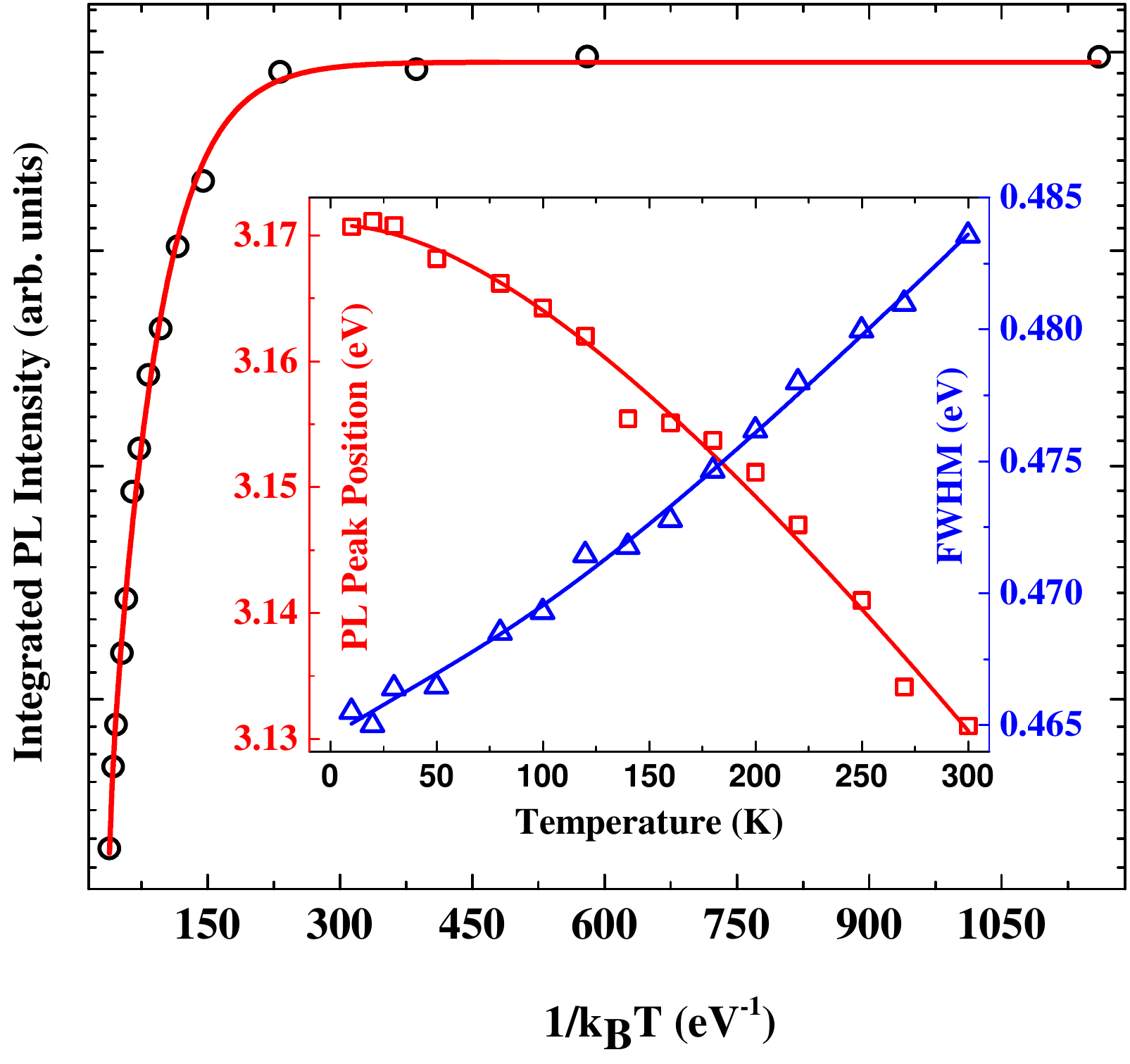}
\caption{\label{TDPL_CdS2_PP_FWHM_II} \footnotesize{Temperature-dependent integrated PL 
intensity corresponding to main resonance energy peak, marked as 2, of CdS2 QD sample. 
The inset shows temperature-dependent PL peak position shift and 
FWHM broadening.}}
\end{figure}

Temperature-dependent photoluminescence (TDPL) 
has been performed to see the effect of temperature 
in the photoluminescence properties of these 
metal / semiconductor composite structures. 
These experiments are performed on spin coated samples. 
P-type (111) silicon wafer is used as substrate. 
The substrates are cleaned using piranha solution, 
a mixture of sulfuric acid ($H_2SO_4$) and hydrogen peroxide ($H_2O_2$). This technique is 
used to clean organic residues off substrates. 
Because the mixture is a strong oxidizing agent, it removes most organic matter, 
and it also hydroxylates most surfaces (add OH groups), 
making them highly hydrophilic (water compatible).
The room temperature PL spectrum profile gets slightly 
modified when the experiments are done on the solid thin films. 
For uncoated solid samples, the peak intensity of the blue luminescence gets slightly
enhanced. Whereas, in case of silver-coated samples, it becomes comparatively stronger than the 
red luminescence. 
This change in PL spectrum profile is attributed to the 
interaction between QDs and silicon substrate.
Furthermore, with changing temperature, 
the PL profile changes drastically.
Fig. \ref{TDPL_CdS2} (a) shows the temperature-dependent 
PL spectra of CdS2 QD sample and the 
fitting of the PL spectrum at 10 K with six best-fit Gaussians is shown in Fig. \ref{TDPL_CdS2} (b). 
The blue PL band corresponding to band-to-band transitions is fitted with three Gaussians as 
the emission spectrum shows two small shoulder peaks 
on either side of the main resonance energy peak marked as 2. 
The high and low energy PL resonance peaks are attributed to the emission from smaller and larger QDs 
present in the sample as shown in Fig. \ref{TEM_CdS_CdS@Ag} (c). 
Whereas, the central PL resonance peak is due to emission from the average size QDs.
The red PL band corresponding to defect state transitions also show two shoulder peaks on either side 
of the main defect PL band marked as 5, 
and therefore fitted with three Gaussians for all temperatures. \\
Different nonradiative processes play an important role in relaxation dynamics of SQDs. 
The processes, responsible for carrier relaxation in SQDs, are radiative relaxation, 
Auger non-radiative scattering \cite{Ghanassi1993}, F\"{o}rster energy transfer between QDs, 
thermal escape from the dot, and trapping in surface and/or defect states \cite{YangPhysRevB1997}. 
Considering the radiative relaxation, a thermally activated 
ionized impurity scattering and the thermal escape of carriers from QDs, 
the temperature dependence of the integrated PL intensity can be given by \cite{Valerini2005}
\begin{equation}
 I_{PL} (T)= \frac{n_0}{1+A e^{-E_a/k_BT}+B(e^{E_{LO}/k_BT}-1)^{-m}},
 \label{IPL_T_final}
\end{equation}
where $n_0$ is the integrated PL intensity at 0 K, $A$ (=$\gamma_{a}/\gamma_{rad}$),
and $B$ (=$\gamma_{0}/\gamma_{rad}$) are the constants related to the 
strength of the involved processes, and $m$ is the number of LO-phonons 
involved in the thermal escape process.
To have a deeper insight into these processes, we analyzed the temperature dependence of the integrated
PL intensity. The experimental data have been fitted to Eq. \ref{IPL_T_final} by fixing 
LO-phonon energy $E_{LO}= 37.2$ meV.
Using activation energy $E_a = 20\pm1$ meV, the best-fit curve 
reproduces the experimental data very well (see Fig. \ref{TDPL_CdS2_PP_FWHM_II}) for $m=5.6\pm0.5$.
Furthermore, as the sample temperature increases, the peak position 
of the central PL band corresponding to band-to-band transitions 
redshifts as shown in the inset of Fig. \ref{TDPL_CdS2_PP_FWHM_II}. 
This can be attributed to temperature-dependent shift of band-gap which is described by empirically 
established Varshni relation \cite{Varshni1967149}: 
\begin{equation}
 E_g (T)=E_g (0)-\alpha\frac{T^2}{T+\beta}, 
\label{Varshni_equation}
\end{equation}
where $E_g (0)$ is the band-gap energy at 0 K, $\alpha$ is the temperature coefficient, 
and $\beta$ is a constant having value close to Debye temperature $\theta_D$ of the material. 
The experimental data has been fitted with Eq. \ref{Varshni_equation}.
The best-fit curve, with $\alpha = (2.6\pm 0.6) 10^{-4}$ eV/K and 
$\beta = 280\pm16$ K, well reproduces the experimental data, though the 
Varshni relation is for bulk semiconductor. The $\beta$ value matches well with the 
theoretically calculated Debye temperature $\theta_D = 300$ K, which is reported in \cite{Yang2012}. 
It demonstrates that the peak position shift 
with temperature is due to the temperature-dependent band-gap shrinkage of CdS, 
while the confinement energies of the carriers are independent of sample temperature. 
Apart from the shift of the spectral position of the peak, it has been observed that the 
FWHM, $\Gamma$, of this peak (inset of Fig. \ref{TDPL_CdS2_PP_FWHM_II}) 
increases with sample temperature. 
Amongst many processes, the exciton-phonon and ionized impurity scattering contribute in modifying FWHM 
with sample temperature.   
The line width broadening with increasing sample temperature in case of semiconductor QDs can be 
explained by the equation \cite{JLeePhysRevB1986},
\begin{equation}
 \Gamma (T) = \Gamma_{inh}+\sigma T +\Gamma_{LO} (e^{E_{LO}/k_BT}-1)^{-1}+
 \Gamma_{imp} e^{-E_a/k_BT},
 \label{Gamma}
\end{equation}
where $\Gamma_{inh}$ is the temperature-independent inhomogeneous broadening which represents 
fluctuations in size, shape, composition, etc. of the QDs. The second and third 
terms are related to homogeneous broadening due to exciton-phonon coupling. $\sigma$ is the 
exciton-acoustic phonon coupling coefficient and $\Gamma_{LO}$ represents exciton-LO phonon 
coupling strength. $E_{LO}$ is the LO-phonon energy and $k_B$ is the Boltzmann constant.
The fourth term of the Eq. (\ref{Gamma}) is related to FWHM broadening due to 
ionized impurity scattering. $E_a$ is the activation energy, required for ionization. 
The experimental data have been reproduced using the first three terms of Eq. \ref{Gamma}. 
The best-fit is obtained for $\Gamma_{inh} = 464.6 \pm 0.3$ meV, 
$\sigma = 48 \pm 4$ $\mu$eV/K, and $\Gamma_{LO} = 15\pm3$ meV. 
The LO-phonon energy $E_{LO} = 37.2$ meV is obtained from the 
Raman spectrum and kept fixed for the fitting. 
Thus, it can be concluded that the ionized impurity scattering is not significant for line-width 
broadening in these CdS QD samples.\par
\begin{figure}[H]
\centering
  \includegraphics[height=0.27\textheight]{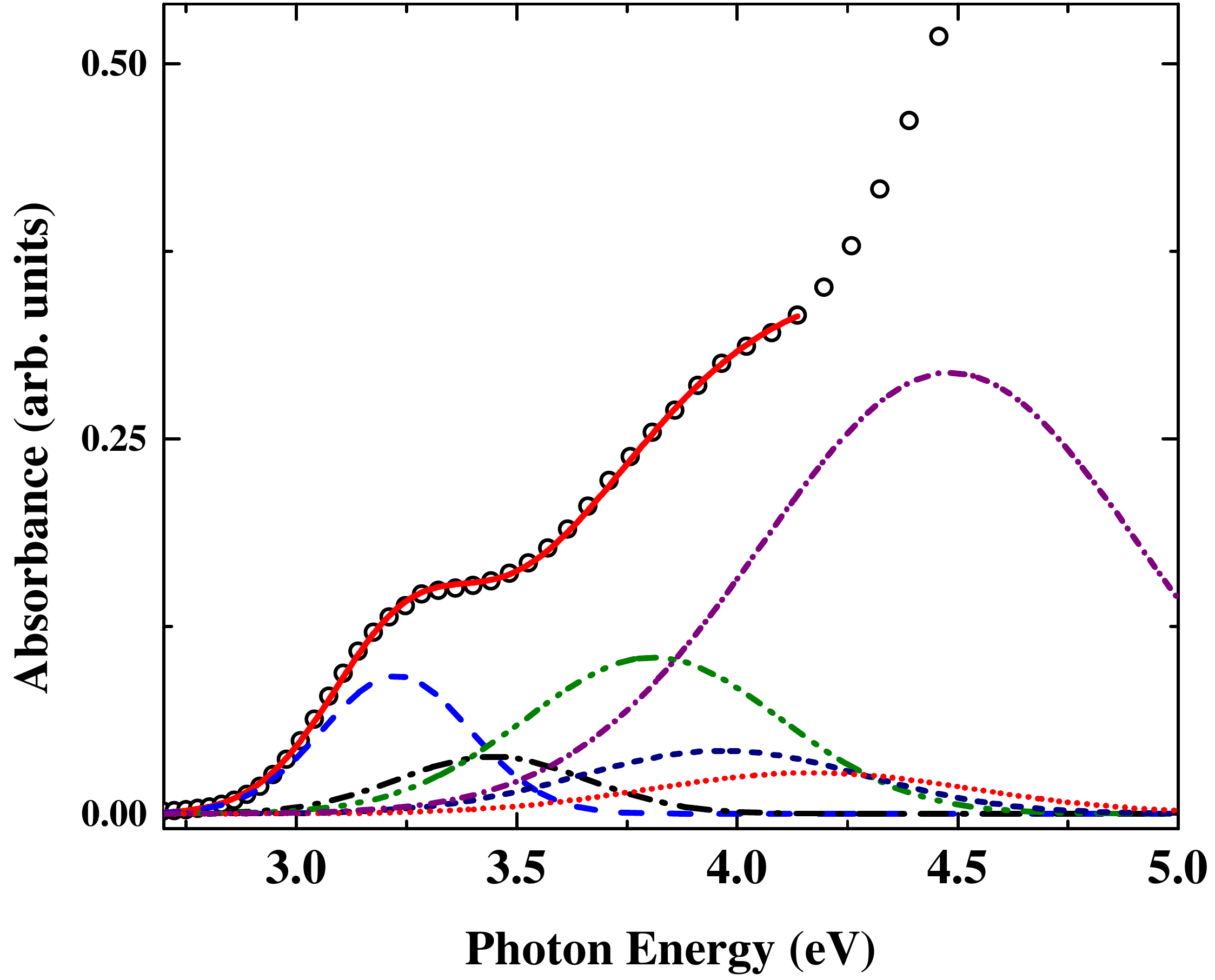}
\caption{\label{TDPL_CdS2_Abs_fitting} \footnotesize{Absorption spectrum of colloidal solution of CdS QDs (CdS2). 
The dot-dashed lines are the best-fit curves of the first six excited states. The continuous curve is 
the cumulative fit.}}
\end{figure}
\begin{figure}[H]
\centering
  \includegraphics[width=0.9\textwidth]{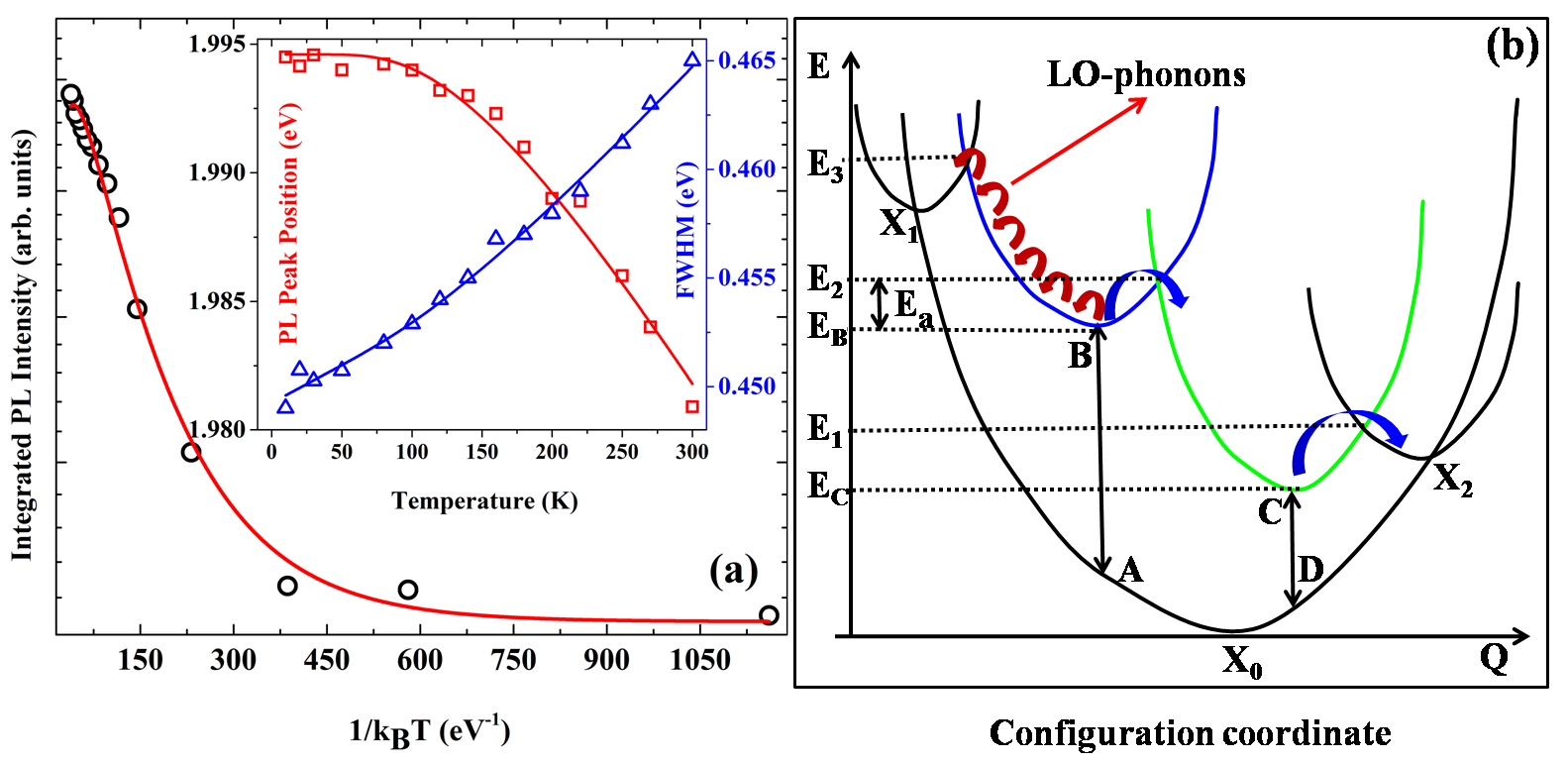}
\caption{\label{TDPL_CdS2_PP_FWHM_II_DS_Eleectronic_States} \footnotesize{
(a) Temperature-dependent integrated PL 
intensity corresponding to main defect PL band for CdS2 QD sample. 
(b) Schematic diagram of configuration coordinate 
geometry showing electronic states of CdS2 QD sample. The blue and green electronic states represent the 
band-edge and defect states, respectively.
The vertical arrows (AB and CD) describe radiative electronic transitions.}}
\end{figure}

In order to understand the origin of the $m$ value, we estimated the energy value of 
the first six excited states from the absorption spectrum of the colloidal solution of CdS QDs. 
The fitting of the absorption spectrum with six Gaussians is shown in Fig. \ref{TDPL_CdS2_Abs_fitting}.
Inhomogeneous broadening $\Gamma_i$ of each excited peak is related to the 
peak energy $E_i=\hbar \omega_i$ and bulk band-gap energy $E_g$ by the 
following equation \cite{Valerini2005, Bawendiprb1996}; 
\begin{equation}
 \Gamma_i=2\delta_R(\hbar \omega_i - E_g),
\end{equation}
where $\delta_R$ is relative size dispersion.
The energy difference between the first two absorption resonance states is around 218 meV, 
which is close to the energy of 6$E_{LO}$. 
Therefore, it can be concluded that the thermal 
escape involves scattering of six LO-phonons. \\
The experimental defect band spectrum has also been reproduced using 
three best-fit Gaussians as shown in Fig. \ref{TDPL_CdS2} (b). 
Fig. \ref{TDPL_CdS2_PP_FWHM_II_DS_Eleectronic_States}  shows temperature-dependent 
integrated PL intensity of the main defect band (peak position $\sim 2$ eV) for CdS QDs.
The fitting of temperature-dependent integrated PL intensity is 
obtained following equation \cite{Jingjpcc2009}, 
\begin{equation}
 I (T) = \frac{n_0+A e^{-E_a/k_BT}}{1+B(e^{E_{LO}/k_BT}-1)^{-m}},
 \label{Integrated_PL_Intensity_Jingjpcc2009}
\end{equation}
where $E_a$ is the activation energy required to overcome the potential barrier, 
A is a parameter that is related to the density of the excited states. 
The best-fit curve reproduces the 
experimental data points very well for the same activation energy as obtained before. \\
The inset of Fig. \ref{TDPL_CdS2_PP_FWHM_II_DS_Eleectronic_States} (a) shows the temperature 
dependent PL peak position shift and 
FWHM broadening of the main defect band (peak position $\sim 2$ eV) for CdS QDs. 
The fitting of temperature dependence of PL peak position is obtained following equation \cite{HsuAPL2007}, 
\begin{equation}
 E (T) = E_0 - \frac{\alpha_0}{e^{E_{LO}/k_BT}-1},
\end{equation}
where $E_{LO}= 37.2$ is LO-phonon energy corresponding to exciton-LO phonon coupling in CdS QD sample. 
We get best-fit curve for $\alpha_0 = 41\pm1$ meV.
The temperature-dependent FWHM of the main defect band for CdS QDs is shown the inset of 
Fig. \ref{TDPL_CdS2_PP_FWHM_II_DS_Eleectronic_States} (a). 
The fitting of temperature-dependent FWHM broadening is obtained following Eq. \ref{Gamma}. 
The best-fit curve reproduces the experimental data points very well 
for $\Gamma_{inh}=449.2\pm0.3$, $\sigma=35\pm4$ $\mu$eV, and $\Gamma_{LO}=16\pm4$ meV.
The different temperature-dependent integrated PL intensity behaviour corresponding to 
band-to-band and defect state transitions for CdS2 QD sample is explained using the 
configuration coordinate geometry which describes the competitive 
radiative and nonradiative transitions associated with crossing of the ground and electronic states. 
Various thermally activated processes are shown in 
Fig. \ref{TDPL_CdS2_PP_FWHM_II_DS_Eleectronic_States} (b).
The integrated PL intensity of the band-to-band PL peak decrease with 
increasing temperature (see Fig. \ref{TDPL_CdS2_PP_FWHM_II}) due to 
scattering of electrons from the luminating state to some higher energetic dark state 
by thermal escape through exciton-LO phonon coupling and thermally activated ionization. 
Whereas, the integrated PL intensity of the 
main defect PL band increases with temperature 
(see Fig. \ref{TDPL_CdS2_PP_FWHM_II_DS_Eleectronic_States} a). 
With increasing sample temperature two competitive processes of carrier generation happen. 
One is thermally activated electrons jump to the 
main defect state from higher energetic excited state, and in the 
other process, electrons get scattered from the main defect band (peak position $\sim 2$ eV) 
to higher energetic position through exciton-LO phonon coupling. 
In this case, with increasing temperature the PL strength of the defect 
band increase suggesting that the first process dominates over the second process.
\subsubsection{TDPL of silver-coated CdS QDs}
\begin{figure}[H]
\centering
  \includegraphics[height=0.32\textheight]{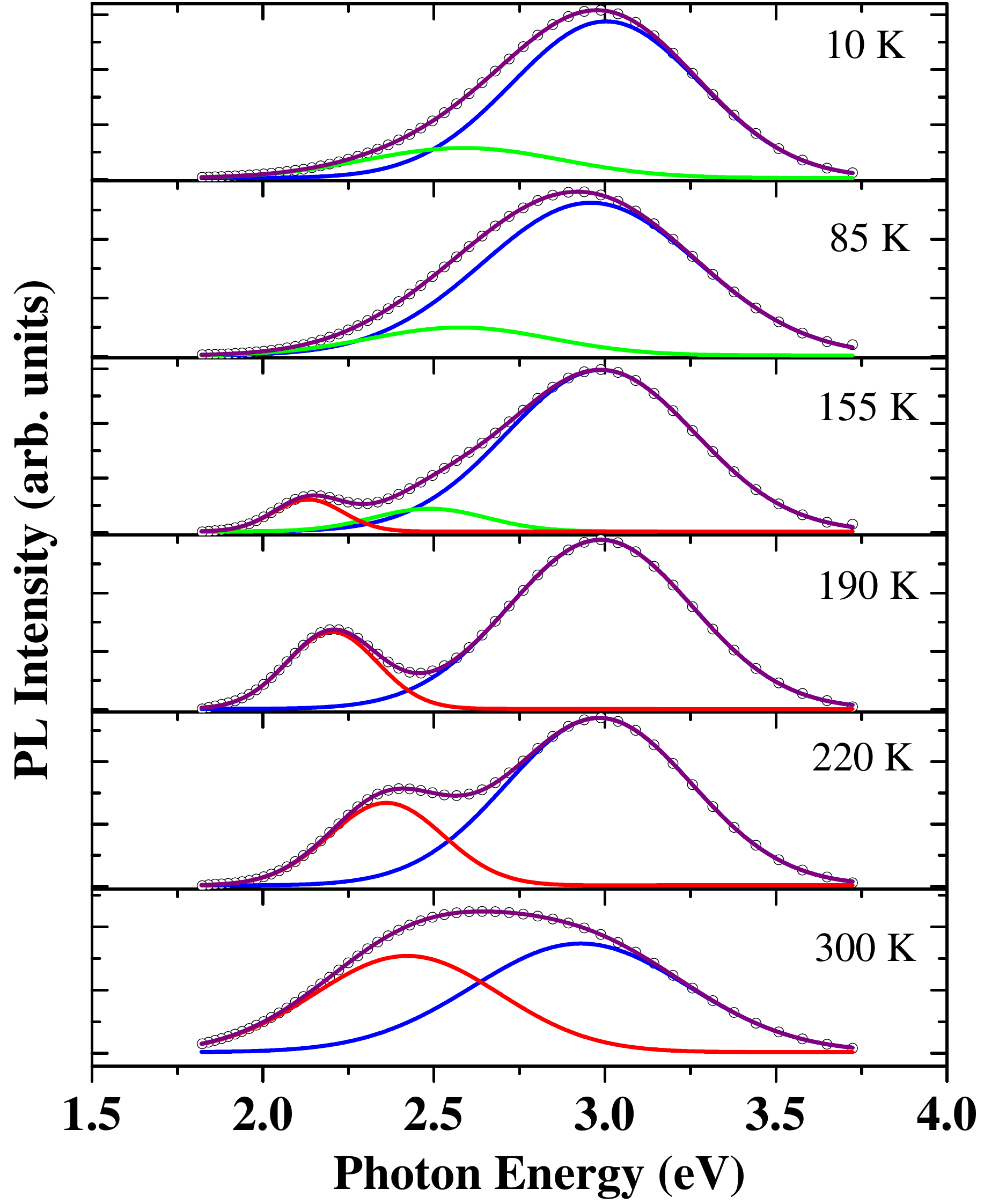}
\caption{\label{TDPL_CdS2@Ag} \footnotesize{Temperature-dependent PL spectra of the 
solid thin film of silver-coated CdS QDs (CdS2@Ag).}}
\end{figure}

\begin{figure}[H]
\centering
  \includegraphics[height=0.3\textheight]{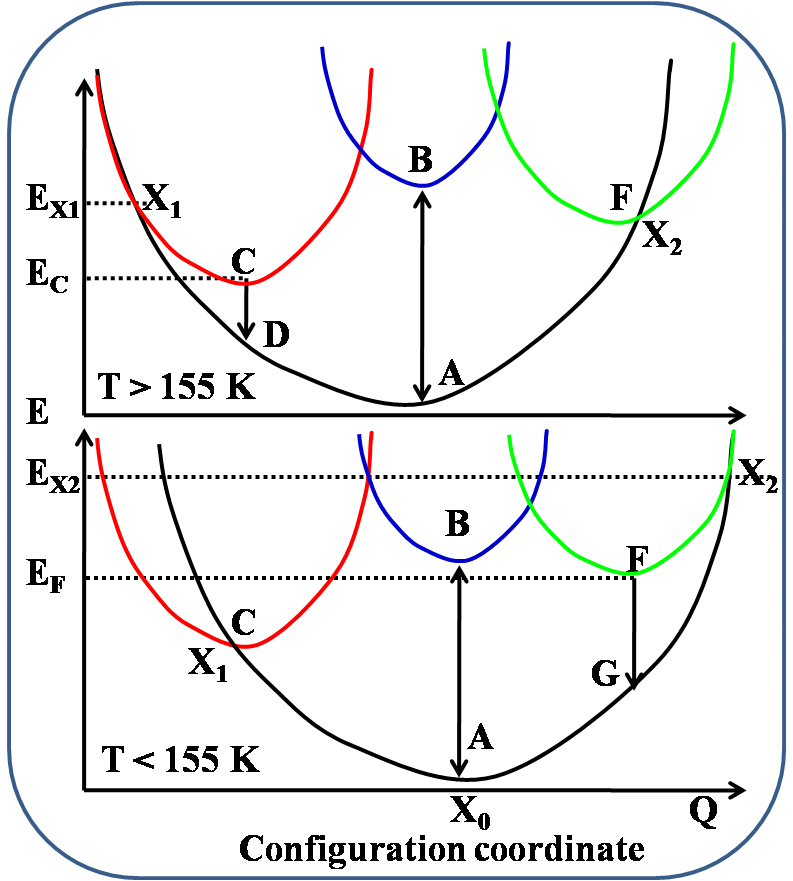}
\caption{\label{TDPL_CdS2@Ag_CC} \footnotesize{Configuration coordinate 
geometry for two different temperature regimes.}}
\end{figure}

\begin{figure}[H]
\centering
  \includegraphics[height=0.3\textheight]{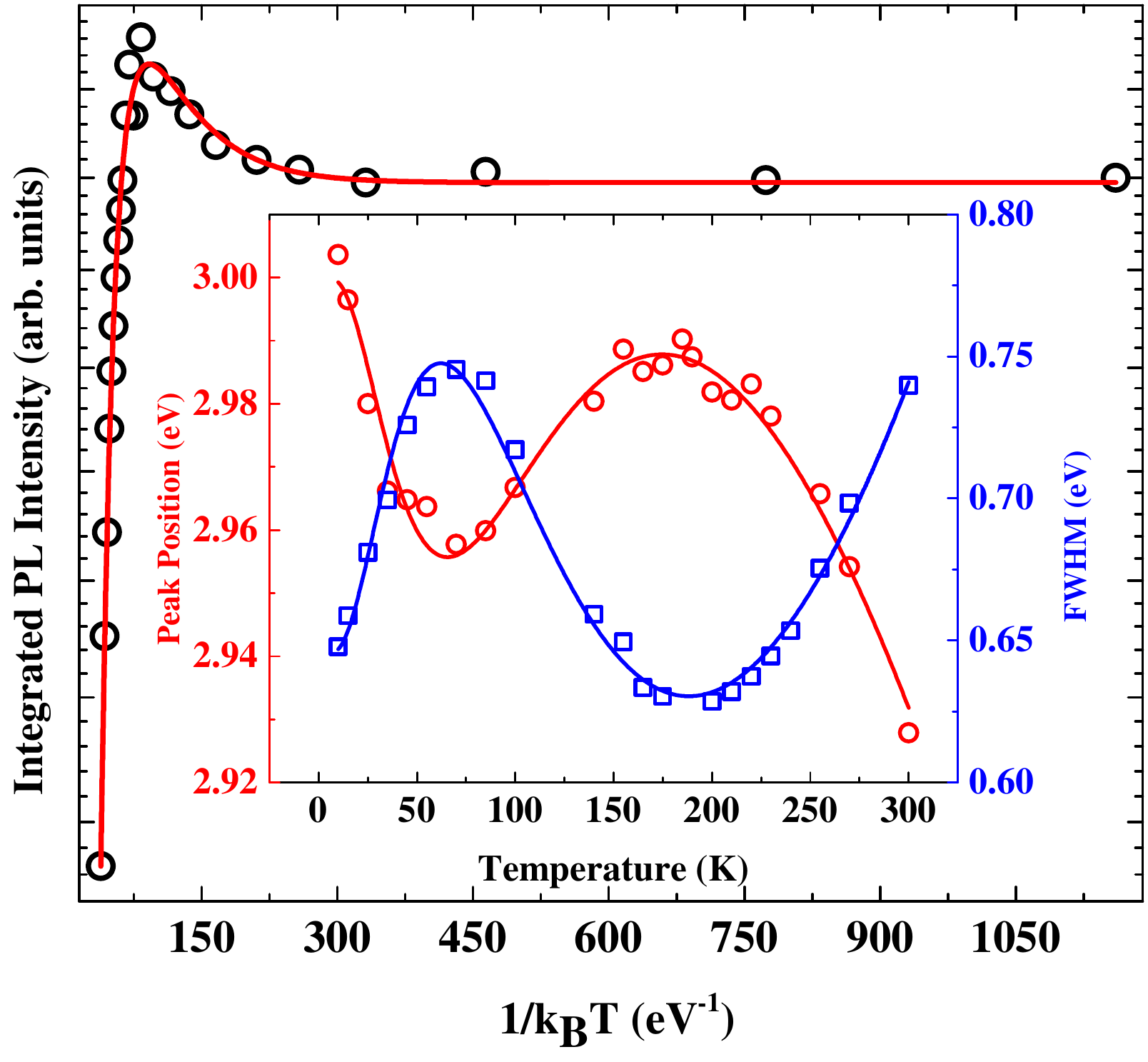}
\caption{\label{TDPL_CdS2@Ag_PP_FWHM_II} \footnotesize{
Observed variation of integrated PL intensity with temperature. 
Inset shows the variation of peak 
position and FWHM with temperature for the band-to-band transition peak for CdS2@Ag sample.}}
\end{figure}

For the sample CdS2@Ag, the temperature-dependent PL 
spectra have been shown in Fig. \ref{TDPL_CdS2@Ag}. 
At room temperature (300 K), the blue luminescence 
band overlaps with the red luminescence band. 
However, as we decrease the temperature, these two PL 
bands become well separated in the temperature 
range 210 K to 155 K. In addition to that, the red PL 
peak shifts further towards the red region of the spectrum 
and the strength of this PL peak goes down and 
does not appear below 140 K. 
In the PL spectrum of this sample, one additional 
structure  appears at 155 K temperature. 
The oscillator strength of this additional 
structure increases as the temperature is 
decreased further. 
This behavior of TDPL spectra of silver-coated CdS QD sample is explained using the 
configuration coordinate geometry for two different temperature regimes. 
Fig. \ref{TDPL_CdS2@Ag_CC} depicts the variation of luminescence efficiency due to 
the change of the relative position of crossover point $X$ with temperature. 
The blue curve is corresponding to the band-edge electronic states, whereas the red and green curves 
represent the defect states. After optical excitation, the $BA$ and $CD$ transitions give 
band-to-band (blue) and defect (red) PL band when the sample temperature is greater than 155 K. 
At this temperature regime, there is no radiative transition possible from the point $F$ to ground 
state as $F$ approaches the crossover point $X_2$. However, as the temperature is lowered, the 
ground state moves towards right with respect to the electronic states. Therefore, the red PL band 
shifts further towards red region of the spectrum and eventually vanishes when the temperature is 
below 155 K as the point $C$ approaches $X_1$ and the electrons cool down non-radiatively. 
Furthermore, at this temperature regime ($T<155$ K), the green PL band starts appearing 
as the crossover point $X_2$ moves away from the point $F$ and initiates radiative transitions ($FG$). \\
Fig. \ref{TDPL_CdS2@Ag_PP_FWHM_II} shows temperature-dependent 
integrated PL intensity of the main resonance PL band (blue) for CdS2@Ag QDs.
The fitting of temperature-dependent integrated PL intensity is obtained following 
Eq. \ref{Integrated_PL_Intensity_Jingjpcc2009}.
The best-fit curve reproduces the experimental 
data points very well for $m=1.9\pm0.2$, $\hbar \omega=37.2$ meV, 
and $\Delta E = 23.9\pm2.4$ meV.
The PL peak energy of the blue band goes through a successive red-blue-red shift (S-shaped) with 
increasing sample temperature as shown in the inset of Fig. \ref{TDPL_CdS2@Ag_PP_FWHM_II}. 
This anomalous temperature-dependent PL characteristic 
can be fitted using a thermal carrier redistribution model \cite{LOURENCO2007}. 
The fitting of temperature dependence of PL peak position of the blue 
PL band is obtained following equation, 
\begin{equation}
 E (T) = E_0 - \frac{A}{e^{\hbar \omega/k_BT}-1}+\frac{B}{e^{\Delta E_1/k_BT}}-\frac{C}{e^{\Delta E_2/k_BT}},
\end{equation}
where $\hbar \omega$ is LO-phonon energy corresponding to exciton-LO phonon coupling and $\Delta E$ 
is activation energy required for ionization. 
We get best-fit curve for the same 
value of $\hbar \omega$ as obtained 
from the temperature-dependent integrated intensity data fitting. 
Whereas, $\Delta E_1$ and $\Delta E_2$ values come out to be $4.3\pm0.3$ meV and $23.9\pm 2.4$ meV.
Inset of Fig. \ref{TDPL_CdS2@Ag_PP_FWHM_II} also shows the temperature-dependent 
FWHM of the main PL band corresponding to band-to-band transitions for CdS2@Ag QDs. 
It shows a successive blue-red-blue shift (inverted S-shaped) with increasing sample temperature.
The fitting of temperature-dependent FWHM broadening is obtained following equation, 
\begin{equation}
\Gamma (T) = \Gamma_{inh} + \frac{A}{e^{\hbar \omega/k_BT}-1}
-\frac{B}{e^{\Delta E_1/k_BT}}+\frac{C}{e^{\Delta E_2/k_BT}}.
\end{equation}
The best-fit curve reproduces the experimental data points very well for the same 
value of $\hbar \omega$ and $\Delta E$ as obtained 
from the temperature-dependent peak-position data fitting.
\section{Conclusions}
CdS QDs have been synthesized using $\gamma$-irradiation technique and subsequently silver coating 
of two QD samples have been prepared. 
The HRTEM studies confirmed the presence of planes 
corresponding to hexagonal-wurtzite and cubic zinc-blende 
crystal structures in the CdS QDs. 
The absorption spectra of the uncoated CdS QDs show the 
expected blue-shift of band-edge with respect to bulk CdS, and 
the absorption spectra of silver-coated CdS QDs show a peak at around 3.4 eV corresponding to the 
coupling of localized surface plasmon with CdS-core. 
The PL spectra for coated and uncoated CdS QD samples show two bands corresponding to 
band-to-band and defect state transitions. 
In case of uncoated samples, the PL peak intensity corresponding to 
band-to-band transitions is weaker compared to defect state transitions. 
However, for silver-coated samples, 
the band-to-band PL peak intensity gets enhanced by ten times due to coupling of surface plasmon of silver-shell with 
excitons of CdS-core. 
Finally, we analyzed the radiative and nonradiative processes involved in the 
relaxation of uncoated and silver-coated CdS QDs. 
For CdS QDs, we demonstrated that TDPL is controlled by 
thermally activated carrier localization with an activation energy of about 20 meV 
and thermal escape by scattering with six LO-phonons. 
Whereas, TDPL of silver-coated CdS QDs involves two thermally activated carrier localization states 
with activation energies of about 24 meV and 4 meV. 
It also involves thermal escape by scattering with two LO-phonons.
These results illustrate the properties of the relaxation processes involved in light emission 
from uncoated and silver-coated CdS QDs, and are important for future applications 
in photonics and optoelectronic devices.
\section*{References}
\bibliographystyle{apsrev}
\bibliography{mybibfile}
\end{document}